\documentclass[prc,aps,twocolumn,preprintnumbers,superscriptaddress]{revtex4}

\usepackage{mathrsfs}
\usepackage{amssymb}
\usepackage{graphicx}
\usepackage{dcolumn}
\usepackage{bm}
\usepackage{graphicx}
\usepackage{float}
\usepackage{longtable}
\usepackage{amsmath}
\usepackage{color}
\usepackage{multirow}
\usepackage[ulem=normalem]{changes}
\usepackage{xcolor}
\usepackage{url}

\newfont{\largemi}{cmmi10}
\newfont{\smallmi}{cmmi6}

\draft

\def\eqref#1{Eq.~(\ref{eq:#1})}

\begin{document}

\title{ The nucleon-pair approximation for nuclei from spherical to deformed regions }

\author{G. J. Fu}
\affiliation{School of Physics Science and Engineering, Tongji University, Shanghai 200092, China}

\author{Calvin W. Johnson}
\affiliation{Department of Physics, San Diego State University, 5500 Campanile Drive, San Diego, CA 92182-1233}

\date{\today}

\begin{abstract}

In this paper we model low-lying states of atomic nuclei in the nucleon-pair approximation of the shell model,  using three approaches to select collective nucleon pairs: the generalized seniority scheme, the conjugate gradient method, and the Hartree-Fock approach.
We find the collective pairs obtained from the generalized seniority scheme provides a good description for nearly spherical nuclei, and those from the conjugate gradient method or the Hartree-Fock approach work well for transitional and deformed nuclei.
Our NPA calculations using collective pairs with angular momenta  0, 2, and 4 (denoted by $SDG$ pairs)  reproduce the nuclear shape evolution in the $N=26$ isotones, $^{46}$Ca, $^{48}$Ti, $^{50}$Cr, and $^{52}$Fe, and yield good agreement with
full configuration-interaction calculations of low-lying states in medium-heavy transitional  and deformed nuclei: $^{44-48}$Ti, $^{48}$Cr, $^{50}$Cr, $^{52}$Fe, $^{60-64}$Zn, $^{64,66}$Ge, $^{84}$Mo, and $^{108-112}$Xe.
Finally, using the $SDGI$-pair approximation we  describe low-lying states of $^{112,114}$Ba, cases difficult to reach by conventional configuration-interaction methods.

\end{abstract}


\vspace{0.4in}

\maketitle

\newpage

\section{Introduction}

The atomic nucleus, a quantum many-body system, can display a variety different modes of collective motions. Here we focus on nuclides with even numbers of protons and of neutrons.
A doubly magic or semimagic nucleus is usually spherical in shape, with yrast states described by the (generalized) seniority scheme \cite{Racah,Talmi,Talmi2,Talmi3}.
Open-shell nuclides, that is, away from doubly closed shells, behave like quantum vibrators or rotors.
Rotational motion is well described by the geometric collective model \cite{BM1,BM2} and the Nilsson model \cite{Nilsson}, by assuming the nucleus has a quadrupole deformation in intrinsic states \cite{Rainwater}.
Rotation arising from intrinsic quadrupole deformation can be embedded in a finite harmonic oscillator single-particle basis using Elliott's theory built upon SU(3) symmetry \cite{Elliott58}, providing us with a microscopic description of rotational motion in the context of the spherical shell model (SM).

Low-lying states of deformed nuclei in the medium-mass region, e.g., $^{48}$Cr, are well described by the SM with effective interactions \cite{Caurier}.
Yet the full SM configuration space for heavy-mass nuclides becomes too large to handle.
The hunt for truncation schemes is a key challenge.
An alternative for describing the quadrupole collectivity of low-lying states is the interacting boson model (IBM) \cite{IBM1,IBM2}.
The building blocks of the IBM are bosons with angular momenta 0 and 2, denoted by $s$ and $d$, which represent collective $S$ and $D$ pairs.
The IBM has been a great success in phenomenological description of vibrational and rotational motions \cite{Nomura1,Nomura2}.
Refs. \cite{OAI,GJ95} reported the connection between the IBM and the SM for nearly-spherical vibrational nuclei and $\gamma$-soft nuclei, but such a relation has never been established for well deformed nuclei.

The nucleon-pair approximation (NPA), an efficient truncation scheme of the full SM configuration space \cite{NPA1,NPA2}, adopts the same idea of the IBM but treats collective nucleon (fermion) pairs with good angular momenta as the degrees of freedom.
If all possible pairs are considered, the NPA model space is precisely equivalent to the full SM space;
if, e.g., the building blocks are restricted to $SD$ pairs, the NPA space is reduced to the $SD$-pair truncated space.
The NPA has been extensively used for the description of nearly spherical nuclei; see Ref. \cite{NPAr} for a review.  For example, low-lying states of semi-magic nuclei are well described by one-dimensional nucleon-pair basis states \cite{Cheng};
the $SD$ pairs are responsible for very low-lying states of vibrational open-shell nuclei \cite{v1,v2};
and the overlap between the $SD$-pair wave function and the SM wave function is larger than 0.9 for low-lying states of nearly spherical nuclei \cite{Lei}.

On the other hand, the $SD$-pair approximation is inadequate to naively reproduce the quadrupole collectivity of rotational nuclei.
For example, for the system with nucleon number $N_{\rm p}=N_{\rm n}=6$ in the $pf$ and $sdg$ shells with a pure quadrupole-quadrupole interaction, the moment of inertia and the $E2$ transition strengths calculated by the $SD$-pair approximation is much smaller than those obtained by the full SM calculation \cite{Zhao2000}.
In  recent work we found instead  that  $SDG$ pairs  derived from the Hartree-Fock (HF) Slater determinant provide us  good descriptions of low-lying states of rotational bands \cite{NPAHF}.
Inspired by those results,  we turn here to a systematic study of deformed nuclei in the medium-heavy mass region, using the nucleon-pair approximation.

The paper is organized as follows.
In Sec. II we briefly introduce the framework of the NPA and three approaches (the generalized seniority-based approach, the conjugate gradient approach, and the HF approach) to determine the inner structure of collective pairs.
In Secs. III and IV we  show that while the collective pairs obtained by the generalized seniority-based approach work for nearly spherical nuclei, those by the conjugate gradient or HF approach provide us good descriptions for low-lying states of transitional and rotational nuclei in the $pf$, $1p0f_{5/2}0g_{9/2}$, and $2s1d0g_{7/2}0h_{11/2}$ shells.
In Sec. V we summarize our results.

\section{Framework}

In this paper we use Greek letters $\alpha$, $\beta,\ldots$ to denote SM single-particle states labeled by $n$, $l$, $j$, $j_z$, and we write the creation operator of a nucleon as $\hat{a}_{\alpha}^{\dagger}$.
We use Latin letters $a$, $b,\ldots$ to denote HF single-particle states, and we write the creation operator as $\hat{c}_{a}^{\dagger}$.

\subsection{The NPA basis state}


In the NPA of the shell model, the building blocks are collective nucleon pairs with various spins $J$, which are defined by
\begin{eqnarray}
 {\hat{A}^{(J)^\dag}} &=& \sum_{j_{\alpha} \leq j_{\beta}} y_{J}(j_{\alpha} j_{\beta})  \left( \hat{a}_{j_{\alpha}}^{\dagger} \times \hat{a}_{j_{\beta}}^{\dagger} \right)^{(J)},
\end{eqnarray}
where ${\hat{a}_{j_{\alpha}}}^{\dagger}$ is the creation operator of a valence nucleon on the SM single-particle orbit $j_{\alpha}$, and $y_{J}(j_{\alpha} j_{\beta})$ is the pair-structure coefficient.
For $2N$ valence protons or neutrons, the NPA basis state with total spin $I$ is constructed by $N$ collective pairs coupled successively, i.e.,
\begin{eqnarray} \label{NPAbasis}
| \varphi^{(I)} \rangle  &=& \left( \cdots (( \hat{A}^{({J}_{1})\dag}\times \hat{A}^{({J}_{2})\dag}  )^{(I_2)} \times \hat{A}^{({J}_{3})\dag}  )^{(I_3)}  \right. \nonumber \\
&& \quad
 \left. \times \cdots \times \hat{A}^{({J}_{N})\dag} \right)^{(I)} |0\rangle ,
\end{eqnarray}
where $I_2$, $I_3,\ldots,I_{N-1}$ are intermediate spins.

In the $SD$-pair approximation, for example, the basis state for valence protons or neutrons is constructed by $SD$ pairs in Eq. (\ref{NPAbasis}).
By choosing the intermediate and total spins $\{I_i\}$ in all possible ways, one usually gets an overcomplete basis. From this we select a maximal linearly independent set of states $\{ | \varphi^{(I)}_i \rangle \}$, which can be chosen in several equivalent ways.
For open-shell nuclei, the basis state is coupled by the proton and neutron states, $ | ( \varphi^{(I_{\pi})}_{i_{\pi}} \times \varphi^{(I_{\nu})}_{i_{\pi}} )^{(I)} \rangle $, where $\pi$ and $\nu$ represent valence protons and neutrons, respectively.
The space spanned by these states is called the $SD$-pair subspace.
Matrix elements of overlaps, one-body operators, and two-body operators in the NPA basis were derived based on the Wick theorem of coupled operators \cite{NPA0}.
Then the calculated states are obtained by diagonalizing Hamiltonian matrix in the $SD$-pair subspace.
Recently, Ref. \cite{He} reported a code based on the Wick theorem derived in the $M$ scheme, allowing one to reach much larger subspaces.

\subsection{The pair-structure coefficient}

In previous work, the pair-structure coefficients, $y_{{J}}(j_{\alpha} j_{\beta})$, of the collective pairs were usually determined by the generalized seniority-like (GS) approach \cite{Xu}.
The detailed procedure is as follows.
The coefficients of the $S$ pair are determined by minimizing the expectation value of Hamiltonian in the $S$-pair-condensation state, i.e.,
\begin{eqnarray}\label{psS}
\frac{ \langle (S)^N_{\tau} | \hat{H} | (S)^N_{\tau} \rangle   }{\langle (S)^N_{\tau} | (S)^N_{\tau} \rangle}, \quad  \text{with}~ \tau=\pi ~\text{or}~ \nu.
\end{eqnarray}
where the creation operator of the $S$ pair can be written as
\begin{eqnarray}\label{psS}
\hat{S}^{\dag} =  \sum_{j_{\alpha}} y_{0}(j_{\alpha} j_{\alpha} )  \left( \hat{a}_{j_{\alpha}}^{\dagger} \times \hat{a}_{j_{\alpha}}^{\dagger} \right)^{(0)},
\end{eqnarray}
For the coefficients of non-$S$ pairs, $y_{{J}}(j_{\alpha} j_{\beta})$, we diagonalize the Hamiltonian matrix in the space spanned by the one-broken pair states $[ (\hat{S}^{\dag})^{N-1} \times ( \hat{a}_{j_{\alpha}}^{\dagger} \times \hat{a}_{j_{\beta}}^{\dagger} )^{(J)} ] |0\rangle$, with $j_{\alpha}$ and $j_{\beta}$ running over all possible single-particle orbits.
The yrast-state wave function can be written by
\begin{eqnarray}
\sum_{j_{\alpha} \leq j_{\beta}} y_J(j_{\alpha} j_{\beta})~ [ (\hat{S}^{\dag})^{N-1} \times ( \hat{a}_{j_{\alpha}}^{\dagger} \times \hat{a}_{j_{\beta}}^{\dagger} )^{(J)} ] |0\rangle.
\end{eqnarray}
The above procedure is done for protons and neutrons independently.
Thus proton-neutron correlations,  important for rotational nuclei, are not considered in the GS approach.

In this work, we use another two approaches to determine the pair-structure coefficients for rotational nuclei: the conjugate gradient (CG) method and the Hartree-Fock (HF) approach.
In the CG approach, the pair-structure coefficients of all pairs adopted in the basis are simultaneously optimized by minimizing the ground-state energy in iterative NPA calculations with a given Hamiltonian.
The advantage of the CG is that it yields the numerically optimal solution.

In the HF approach we extract collective pairs from a HF Slater determinant \cite{NPAHF},
 as follows.
We perform an unconstrained HF calculation in the SM single-particle space with a SM interactions~\cite{SHERPA}.
The HF single-particle states from our calculations are sorted by the HF single-particle energies from the smallest to the largest, which can be written as a transformation of the original SM single-particle states:
\begin{eqnarray}
\hat{c}_{a}^{\dagger} = \sum_{\alpha} U_{a \alpha } \hat{a}_{\alpha}^{\dagger}.
\end{eqnarray}
In general, for even-even nuclei we obtain time-reversed partners of HF single-particle orbits.
A Slater determinant for an even number of valence protons (or neutrons) can be written as a pair condensate:
\begin{eqnarray}\label{HF}
\prod_{a=1}^{2N} \hat{c}_{a}^{\dagger} |0\rangle
&=& (N!)^{-1} \left( \hat{c}_{1}^{\dagger}\hat{c}_{2}^{\dagger} + \cdots + \hat{c}_{2N-1}^{\dagger} \hat{c}_{2N}^{\dagger} \right)^{N} |0\rangle   \nonumber \\
&=& (N!)^{-1} \left(  \sum_{  ab} g_{ ab} ~ \hat{c}^{\dagger}_{a} \hat{c}^{\dagger}_{b} \right)^{N} |0\rangle,
\end{eqnarray}
where $g_{12} = g_{34} = \ldots = g_{(2N-1)(2N)} =1 $ and other $g_{ij} =0$; the phase of each pair is arbitrary.
One can project out pairs of good spin from the deformed HF pair,
\begin{eqnarray}
\hat{B}_{MK}^{(J) \dagger} = \sum_{j_{\alpha} \leq j_{\beta}} y_{JK}(j_{\alpha} j_{\beta}) \left( \hat{a}_{j_{\alpha}}^{\dagger} \times \hat{a}_{j_{\beta}}^{\dagger} \right)^{(J)}_{M},
\end{eqnarray}
where
\begin{eqnarray}\label{yabr}
y_{JK}( j_{\alpha} j_{\beta}) = \displaystyle \sum_{abk_{\alpha}k_{\beta}} g_{ab} ( U_{ a\alpha} U_{b\beta}-U_{ b\alpha} U_{a\beta} ) \frac{ C^{JK}_{j_{\alpha} k_{\alpha} j_{\beta} k_{\beta}} }{ 1+ \delta_{j_{\alpha} j_{\beta}} }.
\label{HFamplitude}
\end{eqnarray}
For a given angular momentum $J$, we diagonalize the norm matrix
\begin{eqnarray}\label{yabr}
N_{KK^{\prime}}^{(JM)} = \langle 0 |  {\hat{B}_{MK}^{(J)} }   \hat{B}_{MK^{\prime}}^{(J) \dagger}  | 0 \rangle.
\end{eqnarray}
The number of nonzero eigenvalues is the number of unique pairs, the nonzero eigenvalues are amplitudes of the unique pairs, and the unique pairs are given by the eigenvectors for the nonzero eigenvalues.

Recently Ref. \cite{PCV} proposed a similar approach, the so-called pair condensate variational (PCV)  approach, to determine pair-structure coefficients.
The PCV approach is somewhat the reverse of our CG approach: both optimize pairs by minimizing the energy expectation value, but in our CG approach we optimize pairs of good angular momenta in the NPA basis, while the PCV optimizes a condensate of a general pair, without enforcing good angular momentum, and then afterwards projects out pairs of good angular momenta.
An NPA calculation using PCV pairs provides a good description for the transitional nuclei $^{132-136}$Ba.

\section{pair approximations with the CG approach}

In this section, we focus  on NPA calculations with the GS and CG approaches.

\begin{figure*}
\includegraphics[width = 0.8 \textwidth]{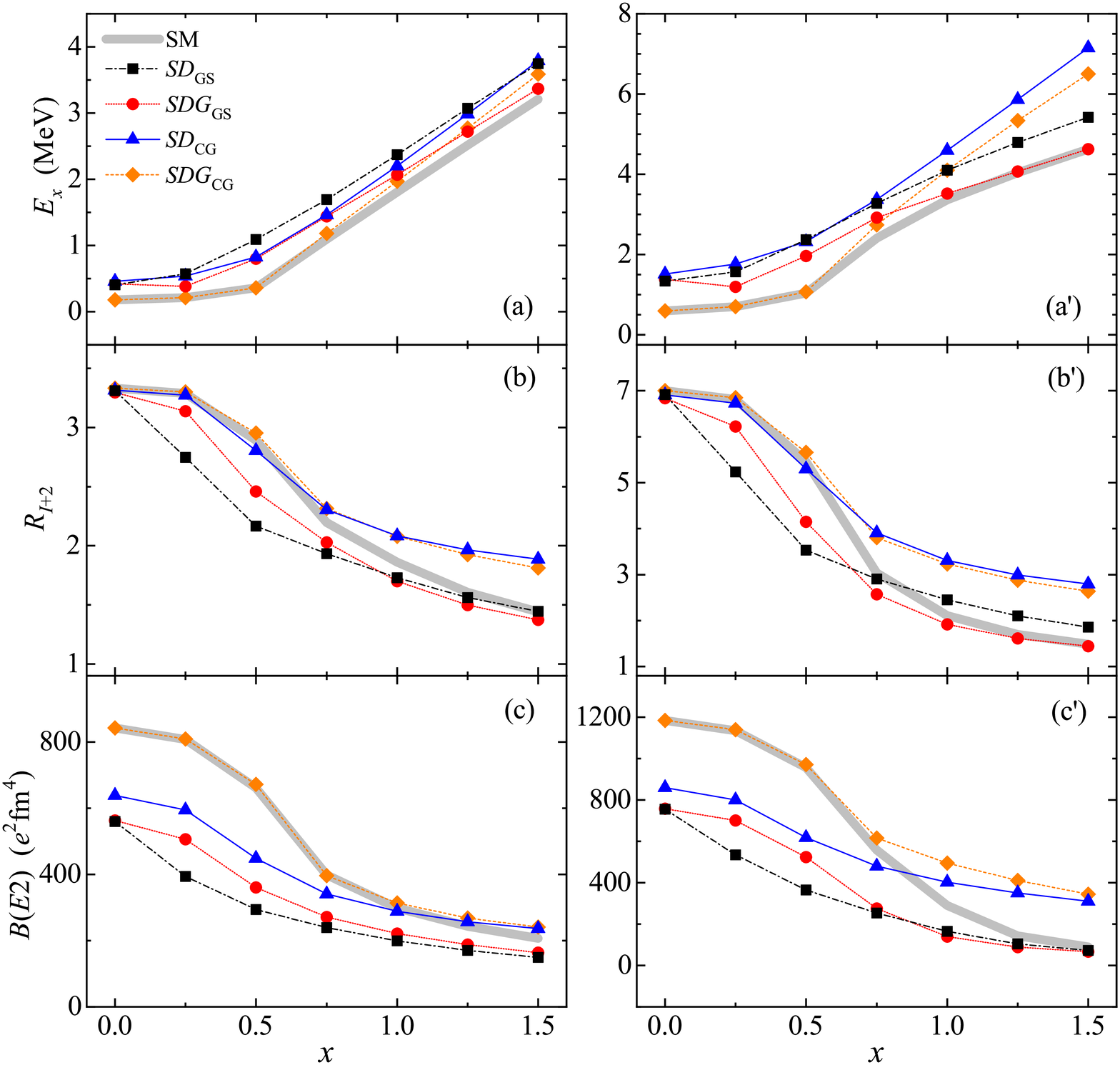}
\caption{\label{fig1} The excitation energies $E_x(I^+_1)$, the energy ratios $R_{I+2}$ (where $R_I \equiv E_x(I^+_1) / E_x(2^+_1)$), and the $B(E2;I \rightarrow I-2)$ values for 6 protons and 6 neutrons in the $pf$ shell with the schematic interaction $H(x)= x \left( \sum_{j_{\alpha}} \varepsilon_{j_{\alpha}}  n_{j_{\alpha}} +   g V_P \right) + \kappa V_Q $ [see Eq. (\ref{H2})].
Panel (a) is for $I=2$ and panel (b) for $I=4$.
SM is the abbreviation for the shell model;
$SDG_{\rm GS}$ is for the $SDG$-pair approximation with the generalized seniority-like approach approach;
and $SDG_{\rm CG}$ is for the $SDG$-pair approximation with the conjugate gradient approach.
}
\end{figure*}

\subsection{Parameter-driven shape evolution}

We investigate the validity of the pair approximation as the nuclear shape evolves from quadrupole deformation to spherical.
Shape evolution can be realized by changing the  ratio of  the strengths of the pairing and  quadrupole-quadrupole interactions
in a schematic Hamiltonian such as
\begin{eqnarray} \label{H2}
H(x) = x \left( \sum_{j_{\alpha}} \varepsilon_{j_{\alpha}}  n_{j_{\alpha}} +   g V_P \right) + \kappa V_Q .
\end{eqnarray}
The first term is the single-particle energy.
The second term,  $V_P$,  is the monopole pairing interaction,
\begin{eqnarray}
&V_P = - {{A}^{(0)}_{\pi}}^{\dag} {A}^{(0)}_{\pi} - {{A}^{(0)}_{\nu}}^{\dag} {A}^{(0)}_{\nu} ,&  \\
&{{A}^{(0)}}^{\dag} = \displaystyle \sum_{j_{\alpha}} \frac{\sqrt{2j_{\alpha}+1}}{2} \left( a_{j_{\alpha}}^{\dagger} \times a_{j_{\alpha}}^{\dagger} \right)^{(0)} .& \nonumber
\end{eqnarray}
The third term $V_Q$ is the quadrupole-quadrupole interaction adopted in the Elliott's SU(3) theory,
\begin{eqnarray}
&V_Q = - (Q_{\pi} + Q_{\nu}) \cdot (Q_{\pi} + Q_{\nu}).&
\end{eqnarray}

This Hamiltonian, or rather, family of Hamiltonians,
we apply in the $pf$ shell with  $N_{\rm p}=N_{\rm n}=6$.
In Eq. (\ref{H2}), the parameters of the single-particle energy are taken from the KB3G effective interaction \cite{kb3g}, i.e., $\varepsilon_{0f_{7/2}} = 0$ MeV,
$\varepsilon_{1p_{3/2}} = 2.0$ MeV,
$\varepsilon_{0f_{5/2}} = 6.5$ MeV,
$\varepsilon_{1p_{1/2}} = 4.0$ MeV;
the strength parameters of the monopole pairing and quadrupole-quadrupole interactions are taken to be $g=0.4$ MeV and $k=0.1$ MeV; and finally
$x$ is an adjustable parameter ranging from 0 to 1.5.

We calculate level energies and the $B(E2)$ transition strength (taking the standard effective charges $e_{\pi}=1.5$ and $e_{\nu}= 0.5$) both in the full SM space using the {\tt BIGSTICK} code \cite{bigstick1,bigstick2}, and in the NPA subspaces.
In Fig. \ref{fig1}  the excitation energies $E_x(I^+_1)$, the energy ratios $R_{I+2}$ (where $R_I \equiv E_x(I^+_1) / E_x(2^+_1)$), and the $B(E2;I \rightarrow I-2)$ values for $I=2$ and 4, all
exhibit evidence of shape evolution, most strongly for the full SM calculations.
For large $x$, that is, large pairing interaction and single-particle splittings, the $2^+_1$ excitation energy is large, accompanied by small ratios $R_4$ and $R_6$, both close to 1.4, and the $B(E2)$ values are small.
These are typical features of spherical nuclei,  well described by the generalized seniority scheme.
As $x$ decreases, the $2^+_1$ and $4^+_1$ excitation energies  decrease rapidly and the energy ratios $R_4$ and $R_6$ as well as the $B(E2)$s increase.
In the case with a dominant quadrupole-quadrupole interaction (where $x$ is close to 0), we find $R_4 \approx 3.33 $ and $R_6 \approx 7$, typical of rotational behavior of deformed nuclei.

While our NPA results follow the SM trends, the details are illuminating.
For large quadrupole-quadrupole interaction (small $x$),
the $SDG$-pair approximation with the CG approach, denoted as $SDG_{\rm CG}$, provides a good description (see in Fig. \ref{fig1}) for the low-lying rotational states:
the excitation energies, energy ratios, and $B(E2)$ values obtained by the $SDG_{\rm CG}$ are in good agreement with the full SM results.
Although the energy ratios in the $SD_{\rm CG}$-pair approximation are close to the SM results, the $SD_{\rm CG}$ underestimates both the $B(E2)$ strengths and the moments
of inertia, i.e., the excitation energies are too large.
The $G$ pair is clearly important in reproducing the collectivity of rotational nuclei.  
The $SDG_{\rm CG}$-pair approximation also provides a good description for the low-lying states of transitional nuclei with $R_4 \sim 2.2$.
The GS approach, conversely, performs poorly for small $x$/strong quadrupole-quadrupole, both in the $SD$- and $SDG$-pair approximations, denoted as $SD_{\rm GS}$ and $SDG_{\rm GS}$,

In the other limiting case, that is, large $x$ / large pairing interaction and single-particle splittings, the results of the $SDG_{\rm CG}$ are not good.
In the CG approach, pair structure coefficients are determined by minimizing the ground-state energy, but
in the limit of a strong pairing interaction, the ground state energy is insensitive to non-$S$ pairs. Thus the CG approach fails to suitably constrain  collective $D$ and $G$ pairs.
While the GS approach is not suitable for the rotational cases, it works rather well for the nearly-spherical cases.

\begin{figure*}
\includegraphics[width = 0.85 \textwidth]{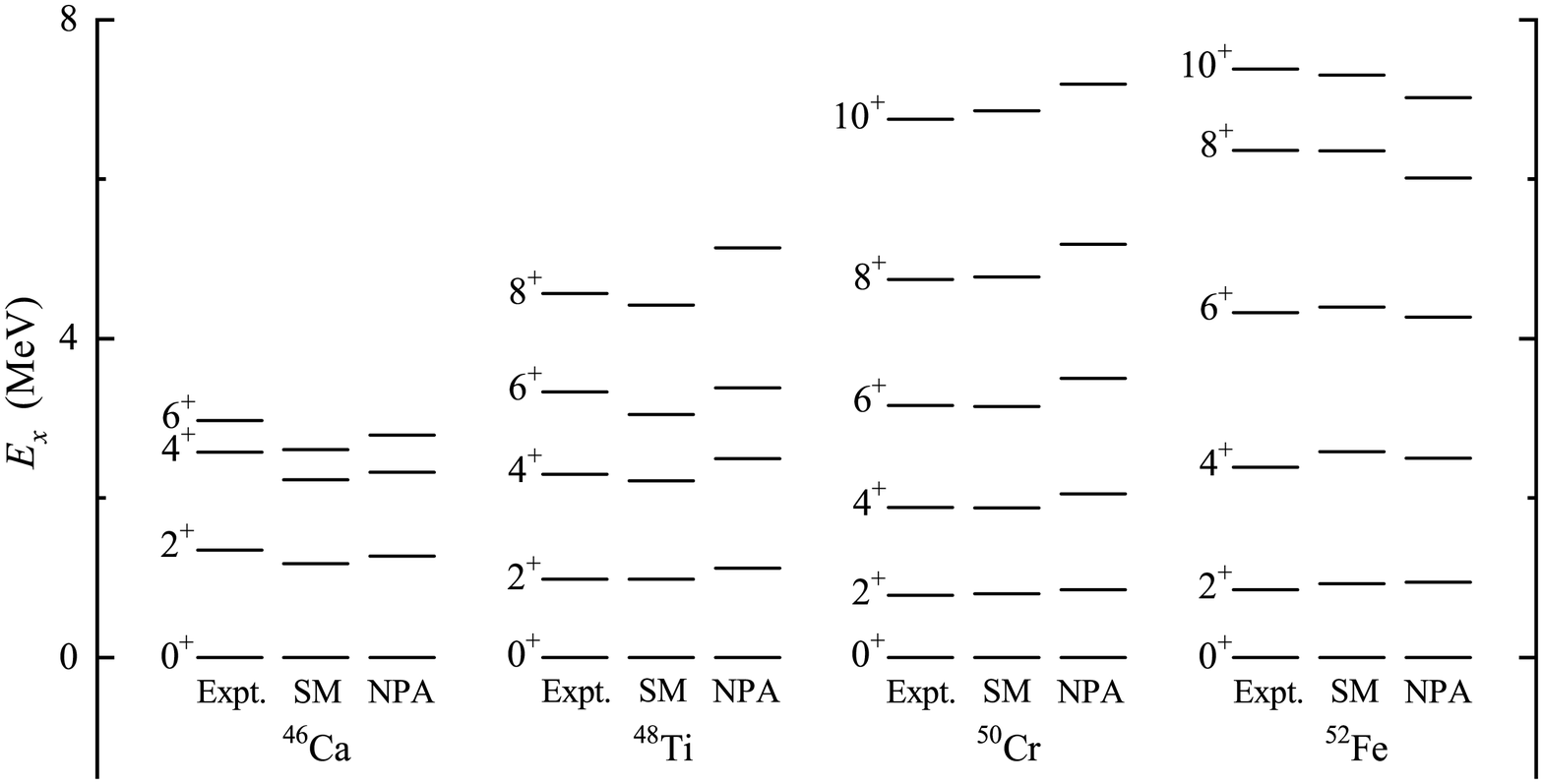}
\caption{\label{fig2} 
	The yrast states of the $N=26$ isotones, $^{46}$Ca, $^{48}$Ti, $^{50}$Cr, and $^{52}$Fe. The NPA calculations are performed in the $SDG_{\rm CG} \bigoplus SDG_{\rm GS}$ space. The experimental data are taken from \cite{NNDC,expt-ca46ti46,expt-ti48cr48,expt-cr50,expt-fe52}.
}
\end{figure*}

\begin{table}
\caption{\label{table1} 
	$B(E2; I \rightarrow I-2)$ (in W.u.) for the yrast states of the $N=26$ isotones, $^{46}$Ca, $^{48}$Ti, $^{50}$Cr, and $^{52}$Fe. The experimental data are taken from \cite{NNDC,expt-ca46ti46,expt-ti48cr48,expt-cr50,expt-fe52,expt-ca46be2}.
    }
 \begin{tabular}{cc|cc|cc|cc|ccc}  \hline\hline
 & Nuclei  &&  $I^{\pi}$  && Expt. && SM &&  NPA &   \\  \hline
 & \multirow{3}{*}{$^{46}$Ca}     && $2^+$ && 2.59(46) && 0.79 && 0.75 &\\
                               &  && $4^+$ && 0.88(21) && 0.65 && 0.64 &\\
                               &  && $6^+$ && 0.55(3) && 0.31 && 0.32 &\\ \hline
 & \multirow{2}{*}{$^{48}$Ti}     && $2^+$ && 14.7(4) && 8.5 && 8.1 &\\
                               &  && $4^+$ && 18.4(17) && 12.7 && 11.4 &\\ \hline
 & \multirow{4}{*}{$^{50}$Cr}     && $2^+$ && 19.3(6) && 16.9 && 15.7 &\\
                               &  && $4^+$ && 14.6(16) && 24.0 && 21.9 &\\
                               &  && $6^+$ && 22(5) && 20.4 && 20.2 &\\
                               &  && $8^+$ && 19(5) && 17.6 && 16.6 &\\ \hline
 & \multirow{4}{*}{$^{52}$Fe}     && $2^+$ && 14.2(19) && 16.0 && 12.5 &\\
                               &  && $4^+$ && 26(6) && 21.3 && 15.6 &\\
                               &  && $6^+$ && 10(3) && 11.8 && 11.2 &\\
                               &  && $8^+$ && 9(4) && 7.2 && 8.3 &\\
       \hline\hline
 \end{tabular}
 \end{table}

\subsection{Nuclear shape evolution in $\bm{N=26}$ isotones }

Continuing our study of shape evolution, we use proton number as a driving parameter.  Experimental data \cite{NNDC,expt-ca46ti46,expt-ti48cr48,expt-cr50,expt-fe52,expt-ca46be2} 
of the low-lying energy levels and the $B(E2)$ values (see in Fig. \ref{fig2} and Table \ref{table1}) show the evolution from spherical to deformed shapes for $N=26$ isotones in the $pf$ shell:
$^{46}$Ca, $^{48}$Ti, $^{50}$Cr, and $^{52}$Fe.
We calculate level energies and $B(E2)$ transition strengths using both the SM and the NPA using a semi-realistic interaction, KB3G \cite{kb3g}.
In the previous subsection, we showed that $SDG_{\rm CG}$ pairs describe well rotational systems and  $SDG_{\rm GS}$ pairs describe well nearly-spherical nuclides.
Thus we expect that NPA calculations in the direct sum of the $SDG_{\rm CG}$ and $SDG_{\rm GS}$ subspaces should reproduce  shape evolution in the $N=26$ isotones (denoted by $SDG_{\rm CG} \bigoplus SDG_{\rm GS}$).

For the yrast states of $^{46}$Ca, $^{48}$Ti, $^{50}$Cr, and $^{52}$Fe,
Fig. \ref{fig2} and Table \ref{table1} compare experimental data and full  SM values to the $SDG_{\rm CG} \bigoplus SDG_{\rm GS}$ results.
Both the level energies and the $B(E2)$ values obtained by the $SDG_{\rm CG} \bigoplus SDG_{\rm GS}$ are in good agreement with experiment and/or full SM results, although both the SM and the $SDG_{\rm CG} \bigoplus SDG_{\rm GS}$
yield  $B(E2; 2^+ \rightarrow 0^+)$ values smaller than experimental for the nearly spherical nuclei $^{46}$Ca and $^{48}$Ti.

\section{pair approximations with the HF approach for nuclei in the medium-heavy mass region}

So far, we have shown that the yrast states of rotational nuclei in the $pf$ shell are well reproduced by the $SDG_{\rm CG}$-pair approximation.
A downside of the CG approach is that hundreds and thousands of iterations of the NPA calculation are needed to reach convergence.
The HF approach is much more practical,  requiring significantly less computing time.  Here we  show that the NPA with pairs from the HF approach provides us with reasonably good description for transitional and deformed nuclei in the medium-heavy mass region.

\begin{figure*}
	\includegraphics[width = 0.92 \textwidth]{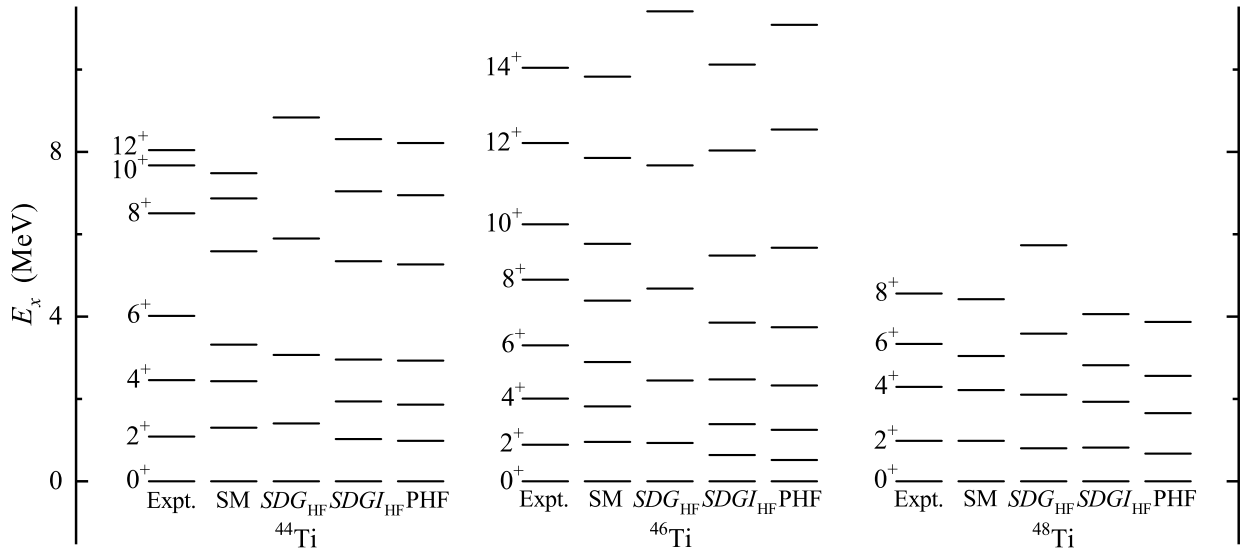}
	\caption{\label{fig3} 
		The yrast states of $^{44-48}$Ti.
		$SDG_{\rm HF}$ is the abbreviation for the $SDG$-pair approximation with the Hartree-Fock approach,
		and PHF is for the angular momentum projected Hartree-Fock calculation. The experimental data are taken from \cite{NNDC,expt-ti44,expt-ca46ti46,expt-ti48cr48}.
	}
\end{figure*}

\begin{table}\center
	\caption{\label{table2} 
		$B(E2; I \rightarrow I-2)$ (in W.u.) for the yrast states of $^{44-48}$Ti. The experimental data are taken from \cite{NNDC,expt-ti44,expt-ca46ti46,expt-ti48cr48}.
	}
	\begin{tabular}{cc|cc|cc|cc|cc|ccc}  \hline\hline
		& Nuclei  &&  $I^{\pi}$  && Expt. && SM &&  $SDG_{\rm HF}$ &&  $SDGI_{\rm HF}$ &   \\  \hline
		& \multirow{4}{*}{$^{44}$Ti}     && $2^+$ && 13(4)    && 12.9 && 13.2 && 13.5 &\\
		&                                && $4^+$ && 30(5)    && 17.0 && 17.6 && 18.6 &\\
		&                                && $6^+$ && 17.0(24) && 14.2 && 15.6 && 18.0 &\\
		&                                && $8^+$ &&          && 10.1 && 11.8 && 16.0 &\\
		\hline
		& \multirow{4}{*}{$^{46}$Ti}     && $2^+$ && 19.5(6)   && 13.2 && 13.7 && 14.1 &\\
		&                                && $4^+$ && 20.2(13)  && 17.4 && 19.9 && 20.3 &\\
		&                                && $6^+$ && 16.4(15)  && 17.6 && 19.7 && 21.4 &\\
		&                                && $8^+$ && 11.3(14)  && 16.1 && 16.7 && 19.1 &\\
		\hline
		& \multirow{4}{*}{$^{48}$Ti}     && $2^+$ && 14.7(4)  && 10.1 && 10.2 && 10.6 &\\
		&                                && $4^+$ && 18.4(17) && 15.0 && 15.7 && 16.2 &\\
		&                                && $6^+$ &&          && 6.0  && 13.5 && 15.2 &\\
		&                                && $8^+$ &&          && 6.9  && 11.6 && 11.0 &\\
		\hline\hline
	\end{tabular}
\end{table}

\begin{figure*}
\includegraphics[width = 0.85 \textwidth]{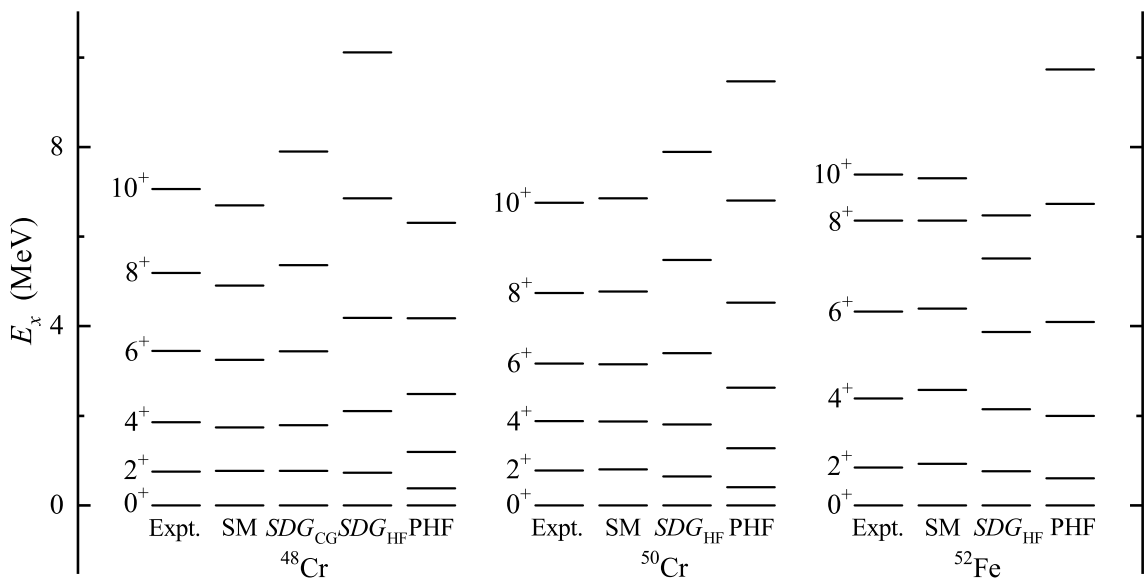}
\caption{\label{fig4} 
	The yrast states of $^{48}$Cr, $^{50}$Cr, and $^{52}$Fe. The experimental data are taken from \cite{NNDC,expt-ti48cr48,expt-cr50,expt-fe52}.
}
\end{figure*}

\begin{table}\center
	\caption{\label{table3} 
		$B(E2; I \rightarrow I-2)$ (in W.u.) for the yrast states of $^{48}$Cr, $^{50}$Cr, and $^{52}$Fe. The experimental data are taken from \cite{NNDC,expt-ti48cr48,expt-cr50,expt-fe52}.
	}
	\begin{tabular}{cc|cc|cc|cc|ccc}  \hline\hline
		& Nuclei  &&  $I^{\pi}$  && Expt. && SM &&  $SDG_{\rm HF}$ &   \\  \hline
		& \multirow{4}{*}{$^{48}$Cr}     && $2^+$ && 31(4) && 20.6 && 19.6 &\\
		&                                && $4^+$ && 27(3) && 28.2 && 27.9 &\\
		&                                && $6^+$ && 29(8) && 28.3 && 29.9 &\\
		&                                && $8^+$ && 24(7) && 26.2 && 29.4 &\\
		\hline
		& \multirow{4}{*}{$^{50}$Cr}     && $2^+$ && 19.3(6) && 16.9 && 18.4 &\\
		&                                && $4^+$ && 14.6(16) && 24.0 && 26.8 &\\
		&                                && $6^+$ && 22(5) && 20.4 && 28.1 &\\
		&                                && $8^+$ && 19(5) && 17.6 && 26.9 &\\
		\hline
		& \multirow{4}{*}{$^{52}$Fe}     && $2^+$ && 14.2(19) && 16.0 && 12.9 &\\
		&                                && $4^+$ && 26(6) && 21.3 && 16.6 &\\
		&                                && $6^+$ && 10(3) && 11.8 && 13.4 &\\
		&                                && $8^+$ && 9(4) && 7.2 && 9.0 &\\
		\hline\hline
	\end{tabular}
\end{table}

\subsection{$\bm{^{44-48}}$Ti, $\bm{^{48}}$Cr, $\bm{^{50}}$Cr, $\bm{^{52}}$Fe in the $\bm{pf}$ shell}

$^{44-48}$Ti are transitional nuclei and $^{48}$Cr, $^{50}$Cr, and $^{52}$Fe are typical deformed nuclei in the $pf$ shell.
For our calculations we use the effective KB3G interaction \cite{kb3g} and for $B(E2)$s take the standard effective charges $e_{\pi}=1.5$ and $e_{\nu}= 0.5$.
The collectivity of the low-lying states are well reproduced by SM calculations, except for $^{44}$Ti.

We begin with transitional nuclei, $^{44-48}$Ti.
For each nuclide we perform the unconstrained HF calculation and obtain a HF state with the minimum energy.
From the HF state, we obtain a unique $S$ pair, a unique $D$ pair, a unique $G$ pair, and a unique $I$ pair.
We find that each of the  $SDGI$ pairs extracted from HF have comparable  amplitudes, that is, the $y_{JK}$ in Eq.~(\ref{HFamplitude}), for these nuclides.
Thus in our NPA calculations, we work in two model spaces:  the $SDG_{\rm HF}$ space, constructed from the unique $SDG$ pairs;  and the $SDGI_{\rm HF}$ space, constructed
using the unique $SDGI$ pairs, with an additional constraint that we allow at most one $I$ pair in our wave functions.
For purposes of comparison, we also compute the excitation energy from  angular momentum projected HF (PHF), using a previously developed method with linear algebra \cite{LAMP}.

Fig. \ref{fig3} and Table \ref{table2} compare for the yrast states of $^{44-48}$Ti values from  experiment \cite{NNDC,expt-ti44,expt-ca46ti46,expt-ti48cr48}, full SM calculations, and the $SDG_{\rm HF}$- and $SDGI_{\rm HF}$-pair approximations.
 Energy levels calculated by  PHF are also included in Fig. \ref{fig3} .
The low-lying spectrum is well reproduced in the SM, except for $^{44}$Ti.
Both of the $SDGI_{\rm HF}$ and the PHF results are in good agreement with the data or the SM results.
The level energies obtained by the $SDG_{\rm HF}$-pair approximation are not good, but interestingly we find that the $B(E2)$ values obtained by the $SDG_{\rm HF}$ are in good agreement with the data or the SM results.

Next we turn to deformed nuclei.
Our unconstrained HF calculation produces an axially symmetric deformed minimum for $^{48}$Cr, $^{50}$Cr, and $^{52}$Fe with $\langle \beta \rangle = $ 0.31, 0.22, and 0.16, respectively.
From each of the HF states, we obtain a unique $S$ pair, a unique $D$ pair, a unique $G$ pair, and a unique $I$ pair, with
the amplitudes of the $SDG$ pairs in the HF state relatively large and that of the $I$ pair significantly  smaller.
Thus in our NPA calculations, we construct our model space, denoted as $SDG_{\rm HF}$, using only $SDG$ pairs.

We compare the experimental data and the SM results against results from the $SDG_{\rm HF}$-pair approximation in
Fig. \ref{fig4} (excitation energies) and in Table \ref{table3} ($B(E2)$ values) for the ground state rotational band.
The energy levels calculated by the PHF are also included in Fig. \ref{fig4}.
Since the configuration space of the $SDG_{\rm HF}$ is much larger than the PHF, one would expect the $SDG_{\rm HF}$ provides us with better results.
Indeed, for $^{50}$Cr and $^{52}$Fe, both the level energies and the $B(E2)$ values obtained by the $SDG_{\rm HF}$ are in remarkably good agreement with the data and the SM results.

For $^{48}$Cr, the $B(E2)$ values and the excitation energies of the yrast $2^+$, $4^+$, and $6^+$ states obtained by the $SDG_{\rm HF}$ are good.
But the excitation energies of the higher-spin $8^+$ and $10^+$ states display an increasing discrepancy.
For comparison, we calculate $^{48}$Cr using the $SDG_{\rm CG}$-pair approximation.  Fig. \ref{fig4} shows
 the $SDG_{\rm CG}$ results (including the higher-spin $6^+$, $8^+$, and $10^+$ states) are closer to the data and the SM results.
The $SD$ pairs from the HF are almost identical to those from the CG (with an overlap larger than 0.99), while the $G$
pairs extracted from the HF and CG approaches differ:
\begin{eqnarray}
\hat{G}^{\dagger}_{\rm HF} &\approx &
0.44 ( \hat{a}_{f_{7/2}}^{\dagger} \times \hat{a}_{f_{7/2}}^{\dagger} )^{(4)}
- 0.61 ( \hat{a}_{f_{5/2}}^{\dagger} \times \hat{a}_{f_{7/2}}^{\dagger} )^{(4)}  \nonumber \\
&& - 0.43 ( \hat{a}_{p_{1/2}}^{\dagger} \times \hat{a}_{f_{7/2}}^{\dagger} )^{(4)}  ,
\end{eqnarray}
\begin{eqnarray}
\hat{G}^{\dagger}_{\rm CG} &\approx &
0.55 ( \hat{a}_{f_{7/2}}^{\dagger} \times \hat{a}_{f_{7/2}}^{\dagger} )^{(4)}
- 0.40 ( \hat{a}_{f_{5/2}}^{\dagger} \times \hat{a}_{f_{7/2}}^{\dagger} )^{(4)}  \nonumber \\
&& - 0.38 ( \hat{a}_{p_{1/2}}^{\dagger} \times \hat{a}_{f_{7/2}}^{\dagger} )^{(4)} .
\end{eqnarray}


\begin{figure*}
	\includegraphics[width = 0.7 \textwidth]{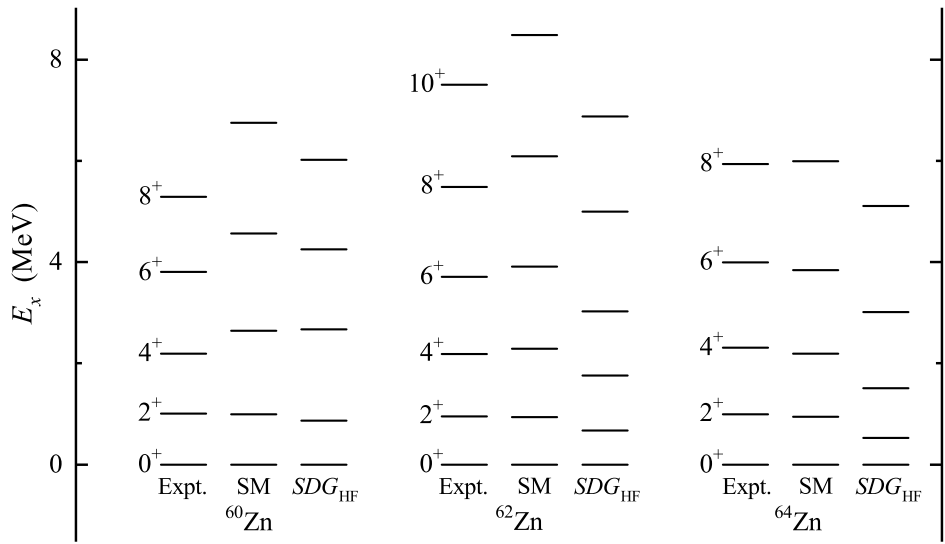}
	\caption{\label{fig5} The yrast states of $^{60-64}$Zn. The experimental data are taken from \cite{NNDC,expt-zn60,expt-zn62,expt-zn64ge64}.
	}	
\end{figure*}

\begin{table}\center
	\caption{\label{table4} $B(E2; I \rightarrow I-2)$ (in W.u.) for the yrast states of $^{60-64}$Zn. The experimental data are taken from \cite{NNDC,expt-zn62,expt-zn64ge64,expt-zn62be2}.
	}
	\begin{tabular}{cc|cc|cc|cc|ccc}  \hline\hline
		& Nuclei  &&  $I^{\pi}$  && Expt. && SM &&  $SDG_{\rm HF}$ &   \\  \hline
		& \multirow{4}{*}{$^{60}$Zn}          && $2^+$ &&   && 18.4  && 18.1   &\\
		&                                     && $4^+$ &&   && 21.6  && 21.8   &\\
		&                                     && $6^+$ &&   && 20.1  && 17.9   &\\
		&                                     && $8^+$ &&   && 12.5  && 9.8   &\\
		\hline
		& \multirow{4}{*}{$^{62}$Zn}          && $2^+$ && 16.8(8)        && 21.0  &&  22.2  &\\
		&                                     && $4^+$ && 26($+7-12$)    && 20.1  &&  22.6  &\\
		&                                     && $6^+$ && 19(3)          && 27.5  &&  25.7  &\\
		&                                     && $8^+$ && 7.9($+20-40$)  && 21.4  &&  21.2  &\\
		\hline
		& \multirow{4}{*}{$^{64}$Zn}          && $2^+$ && 20.0(6)  && 21.0  && 23.4   &\\
		&                                     && $4^+$ && 12.2(5)  && 25.8  && 31.1   &\\
		&                                     && $6^+$ && 23(6)    && 24.9  && 28.7   &\\
		&                                     && $8^+$ &&          && 15.5  && 20.7   &\\
		\hline

		\hline\hline
	\end{tabular}
\end{table}

\begin{figure*}
\includegraphics[width = 0.8 \textwidth]{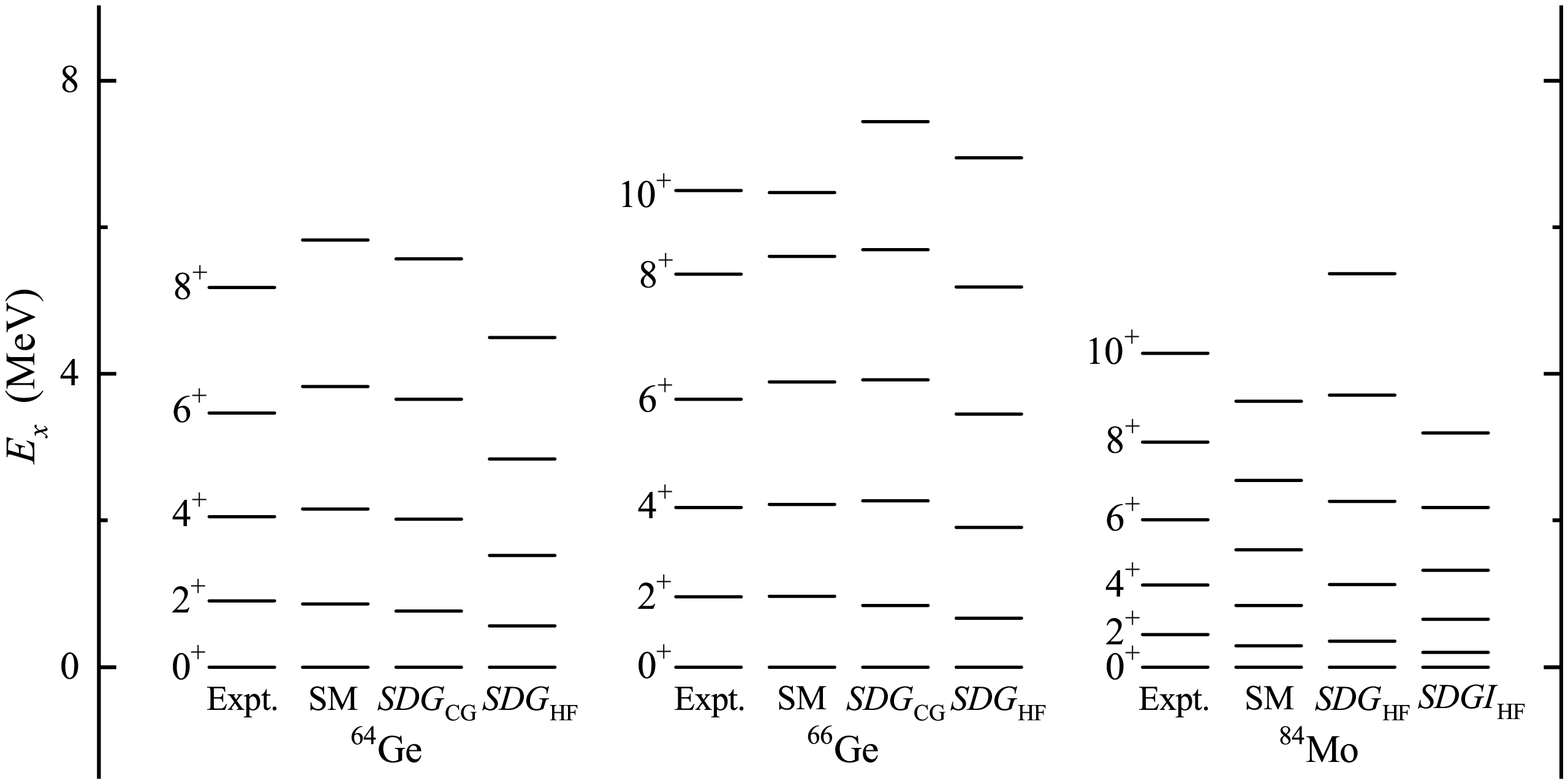}
\caption{\label{fig6} The yrast band of $^{64,66}$Ge and $^{84}$Mo. The experimental data are taken from \cite{NNDC,expt-zn64ge64,expt-ge66,expt-mo84}.
}	
\end{figure*}

\subsection{$\bm{^{60-64}}$Zn, $\bm{^{64,66}}$Ge, $\bm{^{84}}$Mo in the $\bm{pf5g9}$ shell}

\begin{table}\center
	\caption{\label{table5} $B(E2; I \rightarrow I-2)$ (in W.u.) for the yrast band of $^{64,66}$Ge, $^{84}$Mo. The experimental data are taken from \cite{NNDC,expt-ge66}.
	}
	\begin{tabular}{cc|cc|cc|cc|cc|ccc}  \hline\hline
		& Nuclei  &&  $I^{\pi}$  && Expt. && SM &&  $SDG_{\rm CG}$ &&  $SDG_{\rm HF}$ &   \\  \hline
		& \multirow{4}{*}{$^{64}$Ge}     && $2^+$ &&  && 28.0 && 27.2 && 29.3 &\\
		&                                     && $4^+$ &&  && 35.0 && 38.6 && 40.2 &\\
		&                                     && $6^+$ &&  && 43.9 && 39.7 && 43.7 &\\
		&                                     && $8^+$ &&  && 36.6 && 30.3 && 41.1 &\\
		\hline
		& \multirow{5}{*}{$^{66}$Ge}     && $2^+$ && 12.0(23) && 28.5 && 24.9 && 24.4 &\\
		&                                     && $4^+$ && $>9.6$   && 32.5 && 33.3 && 32.2 &\\
		&                                     && $6^+$ && $>1.2$   && 34.1 && 32.1 && 30.4 &\\
		&                                     && $8^+$ &&          && 14.8 && 19.0 && 21.4 &\\
		&                                     && $10^+$&& $<5.1$   && 0.1  && 7.1  && 8.1 &\\
		\hline\\
		\hline
		& Nuclei  &&  $I^{\pi}$  && Expt. && SM &&  $SDG_{\rm HF}$ &&  $SDGI_{\rm HF}$ &   \\  \hline
		& \multirow{5}{*}{$^{84}$Mo}          && $2^+$ &&  && 54.9 && 48.8  && 58.2  &\\
		&                                     && $4^+$ &&  && 79.2 && 68.7  && 82.3  &\\
		&                                     && $6^+$ &&  && 86.8 && 73.1  && 88.9  &\\
		&                                     && $8^+$ &&  && 89.1 && 72.5  && 90.5  &\\
		&                                     && $10^+$&&  && 89.1 && 68.9  && 89.5  &\\
		\hline\hline
	\end{tabular}
\end{table}

We calculate
low-lying states of $^{60-64}$Zn, $^{64,66}$Ge and $^{84}$Mo    in the $1p_{1/2}$-$1p_{3/2}$-$0f_{5/2}$-$0g_{9/2}$ space, denoted ${pf5g9}$, using  the JUN45 interaction \cite{JUN45}, which
provides a reasonably good description in the full SM.  
For our NPA calculations we use HF-derived pairs from the same interaction.
The effective charges used for the $B(E2)$ calculation are $e_{\pi}=1.5$ and $e_{\nu}= 1.1$.

The $^{60-64}$Zn isotopes have two valence protons in the ${pf5g9}$ shell, and thus the collectivity is not very strong.
For each of the isotopes, we perform the unconstrained HF calculation and obtain a HF state with the minimum energy.
From the HF state, we obtain a unique $S$ pair, a unique $D$ pair, and a unique $G$ pair.
The amplitude of the $G$ pair is non-negligible, and so in the NPA calculation of $^{60-64}$Zn, we construct our model space using $SDG$ pairs, i.e., the $SDG_{\rm HF}$-pair approximation.
Fig. \ref{fig5} and Table \ref{table4} compare the yrast states of $^{60-64}$Zn from the experimental data \cite{NNDC,expt-zn60,expt-zn62,expt-zn64ge64,expt-zn62be2}, the SM, and the $SDG_{\rm HF}$-pair approximation.
Both the level energies and the $B(E2)$ values obtained by the $SDG_{\rm HF}$ are in good agreement with the data or the SM results.

Our unconstrained HF calculation produces a local minimum with $\langle \beta \rangle = 0.28$ and $\langle \gamma \rangle = 20^{\circ}$ for $^{64}$Ge, which indicates a triaxially deformation, and one with $\langle \beta \rangle = 0.21$ and $\langle \gamma \rangle = 60^{\circ}$ for $^{66}$Ge, which has an axially symmetric oblate deformation.
From the triaxially deformed HF state, we obtain a unique $S$ pair, two different $D$ pairs, and two different $G$ pairs.
The amplitudes of the second $D$ pair and the second $G$ pair are very small.
Therefore, we construct our model space for $^{64}$Ge using the unique $S$ pair and the first $DG$ pairs, i.e., the $SDG_{\rm HF}$-pair approximation.
From the oblate HF state, we obtain a unique $S$ pair, a unique $D$ pair, and a unique $G$ pair, and thus calculate $^{66}$Ge using the $SDG_{\rm HF}$ pairs.
For comparison, we also calculate these two nuclei using the $SDG_{\rm CG}$-pair approximation.

In Fig. \ref{fig6} we see that the low-lying states of $^{64}$Ge and $^{66}$Ge from the data \cite{NNDC,expt-zn64ge64,expt-ge66}, the SM, the $SDG_{\rm CG}$, and the $SDG_{\rm HF}$ are in good agreement with each others.
The $B(E2)$ values obtained by the SM, the $SDG_{\rm CG}$, and the $SDG_{\rm HF}$ are also very close to each others, except for $B(E2;10^+ \rightarrow 8^+)$ (see in Table \ref{table5}).
The above result shows that both axially symmetric deformed nuclei and triaxially deformed nuclei can be well described by using the $SDG_{\rm HF}$- and $SDG_{\rm CG}$-pair approximations.

For $^{84}$Mo in the particle formalism (with 14 valence protons and 14 valence neutrons) our unconstrained HF calculation produces an oblate minimum with $\langle \beta \rangle = 0.13$ and $\langle \gamma \rangle = 60^{\circ}$.
However it is more convenient to carry out the NPA calculation in the hole formalism, with  8 valence proton holes and 8 valence neutron holes, and a Pandya transformation on the interaction.
From the HF state in the hole formalism, we obtain a unique $S$ pair, a unique $D$ pair, a unique $G$ pair, and a unique $I$ pair.
The amplitudes of the $SD$ pairs are large.
Although the amplitudes of the $GI$ pairs are relatively smaller, they are non-negligible.
Thus we calculate $^{84}$Mo in the hole formalism by using the $SDG_{\rm HF}$- and $SDGI_{\rm HF}$-pair approximations (in the latter the maximum numbers of the $GI$ pairs are constrained to one for simplicity).

Fig. \ref{fig6} and Table \ref{table5} compare the excitation energies and $B(E2)$ values between the data \cite{NNDC,expt-mo84}, the SM, the $SDG_{\rm HF}$- and $SDGI_{\rm HF}$-pair approximation results for $^{84}$Mo. The level energies of the low-lying $2^+$ and $4^+$ states obtained by the $SDG_{\rm HF}$ are in good agreement with the SM results, but for higher-spin states we see increasing discrepancies.
The $B(E2)$ values from the $SDG_{\rm HF}$ result are $11$-$23 \%$ smaller than those from the SM result.
The above result suggests the $I$ pair might be important.
Indeed, for the level energies and the $B(E2)$ values, the agreement between the $SDGI_{\rm HF}$ and the SM results are significantly improved, even if the former predict a moment of inertia slightly larger than the latter.

\begin{figure*}
\includegraphics[width = 0.7 \textwidth]{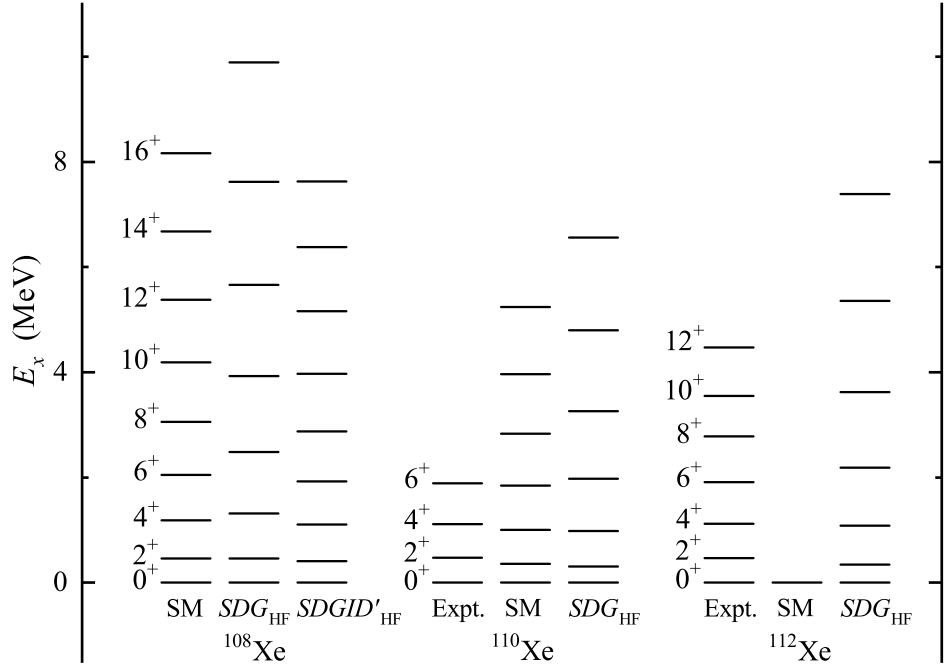}
\caption{\label{fig7} The yrast band of $^{108-112}$Xe. The experimental data are taken from \cite{NNDC,expt-xe110,expt-xe112}.
}
\end{figure*}

\begin{figure}
\includegraphics[width = 0.48 \textwidth]{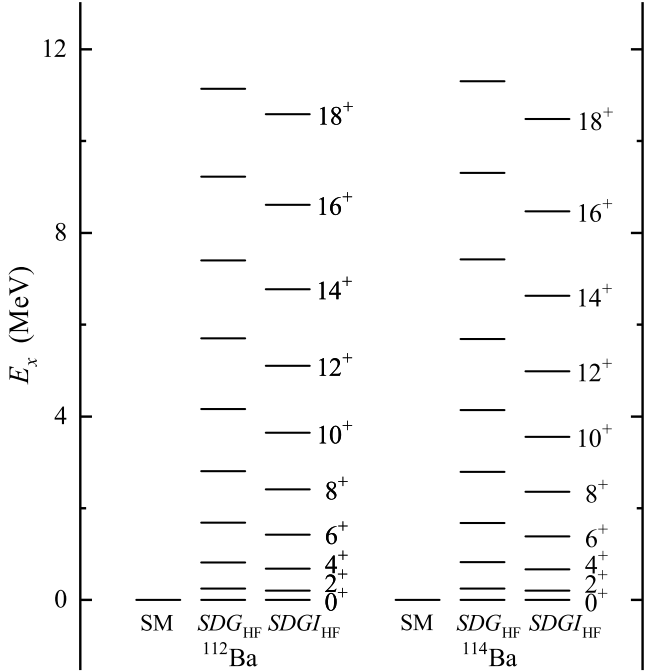}
\caption{\label{fig8} The yrast band of $^{112,114}$Ba.
}
\end{figure}

\begin{table}\center
	\caption{\label{table6} $B(E2; I \rightarrow I-2)$ (in W.u.) for the yrast states of $^{108-112}$Xe and $^{112,114}$Ba.
	}
	\begin{tabular}{cc|cc|cc|cc|ccc}  \hline\hline
		& Nuclei  &&  $I^{\pi}$ && SM &&  $SDG_{\rm HF}$ &&  $SDGID^{\prime}_{\rm HF}$ &   \\  \hline
		& \multirow{5}{*}{$^{108}$Xe}         && $2^+$ && 30.9 && 27.6 && 30.2 &\\
		&                                     && $4^+$ && 41.8 && 38.7 && 40.5 &\\
		&                                     && $6^+$ && 44.9 && 40.8 && 43.1 &\\
		&                                     && $8^+$ && 47.4 && 39.4 && 45.0 &\\
		&                                     && $10^+$&& 44.8 && 35.6 && 43.0 &\\
		\hline\\
	\end{tabular}
    \begin{tabular}{cc|cc|cc|ccc}  \hline
		& Nuclei  &&  $I^{\pi}$  && SM &&  $SDG_{\rm HF}$ &   \\  \hline
		& \multirow{5}{*}{$^{110}$Xe}         && $2^+$ && 34.3 && 33.3 &\\
		&                                     && $4^+$ && 48.4 && 46.6 &\\
		&                                     && $6^+$ && 51.7 && 49.6 &\\
		&                                     && $8^+$ && 52.1 && 49.3 &\\
		&                                     && $10^+$&& 50.8 && 47.1  &\\
		\hline
		& \multirow{5}{*}{$^{112}$Xe}         && $2^+$ &&      && 33.9 &\\
		&                                     && $4^+$ &&      && 48.1 &\\
		&                                     && $6^+$ &&      && 51.8 &\\
		&                                     && $8^+$ &&      && 52.1 &\\
		&                                     && $10^+$&&      && 50.1 &\\
		\hline\\
	\end{tabular}
    \begin{tabular}{cc|cc|cc|cc|ccc}  \hline
		& Nuclei  &&  $I^{\pi}$  && SM &&  $SDG_{\rm HF}$ &&  $SDGI_{\rm HF}$ &   \\  \hline
		& \multirow{5}{*}{$^{112}$Ba}         && $2^+$ &&  && 50.2  && 54.0  &\\
		&                                     && $4^+$ &&  && 70.6  && 76.0  &\\
		&                                     && $6^+$ &&  && 75.5  && 81.3  &\\
		&                                     && $8^+$ &&  && 75.6  && 81.4  &\\
		&                                     && $10^+$&&  && 73.0  && 78.7  &\\
		\hline
		& \multirow{5}{*}{$^{114}$Ba}         && $2^+$ &&  && 50.2  && 54.6  &\\
		&                                     && $4^+$ &&  && 70.6  && 76.9  &\\
		&                                     && $6^+$ &&  && 75.5  && 82.4  &\\
		&                                     && $8^+$ &&  && 75.9  && 82.8  &\\
		&                                     && $10^+$&&  && 73.8  && 80.4  &\\
		\hline\hline
	\end{tabular}
\end{table}

\subsection{$\bm{^{108-112}}$Xe, $\bm{^{112,114}}$Ba in the $\bm{sdg7h11}$ shell}

The lightest Xe and Ba isotopes have been observed to be $^{108}$Xe and $^{112}$Ba recently, which are $N=Z$ nuclei.
Although the low-lying spectra are unknown, the excitation energies of low-lying states of their neighbors $^{110}$Xe and $^{112}$Xe have been measured, which show   collective rotational features \cite{NNDC,expt-xe110,expt-xe112}.
In this work, we calculate $^{108-112}$Xe using both the SM and the NPA with the HF approach in the $2s_{1/2}$-$1d_{3/2}$-$1d_{5/2}$-$0g_{7/2}$-$0h_{11/2}$ space, denoted  ${sdg7h11}$), with the monopole-optimized effective interactions based on the CD-Bonn potential renormalized by the perturbative G-matrix approach \cite{qi}.
We also compute  $^{112,114}$Ba in the NPA with HF-derived pairs using the same single-particle space and interaction. 
The shell-model $M$-scheme dimensions of $^{112}$Xe and $^{112}$Ba, are $\sim$9 billion and $\sim$20 billion, respectively, at the edge of what the modern large-scale SM can do, 
while the $M$-scheme dimension of $^{114}$Ba, is $\sim$220 billion, well beyond the reach of any full SM calculation. 
These nuclei demonstrate the applicability and utility of the NPA.

Our HF calculation of $^{108}$Xe produces a triaxially deformed minimum with $\langle \beta \rangle = 0.39$ and $\langle \gamma \rangle = 11^{\circ}$. From this HF state, we obtain one unique $S$ pair, two $D$ pairs, two $G$ pairs, and two $I$ pairs.
The amplitudes of the second $DGI$ pairs are much smaller than those of the first ones.
Our NPA model space is constructed using the first $SDG$ pairs.
Since the amplitude of the first $I$ pair is non-negligible, and the amplitude of the second $D$ pair (denoted by $D^{\prime}$) are larger than those of the second $GI$ pairs, we also perform an NPA calculation in the space constructed from the first $SDGI$ pairs and the $D^{\prime}$ pair (for simplicity the maximum number of the $D^{\prime}$ pair is constrained to one).
Our HF calculation produces a prolate minimum with $\langle \beta \rangle = 0.39$ and $\langle \gamma \rangle = 0^{\circ}$ for $^{110}$Xe and a triaxially deformed minimum with $\langle \beta \rangle = 0.39$ and $\langle \gamma \rangle = 11^{\circ}$ for $^{112}$Xe.
Our NPA model spaces for these two nuclei are constructed using the first $SDG$ pairs.

Fig. \ref{fig7} and Table \ref{table6} compare low-lying states of the ground rotational band from the data \cite{NNDC,expt-xe110,expt-xe112}, the SM, and the NPA with the HF approach.
The level energies of the low-lying $2^+$, $4^+$, and $6^+$ states obtained by the $SDG_{\rm HF}$ are in good agreement with the data or the SM results, and the same to the $B(E2)$ values for $2^+ \rightarrow 0^+$, $4^+ \rightarrow 2^+$, and $6^+ \rightarrow 4^+$.
But for the higher-spin states of $^{108}$Xe we see increasing discrepancy.
For the level energies and the $B(E2)$ values, the agreement between the $SDGID^{\prime}_{\rm HF}$ and the SM results is significantly improved, suggesting the high-spin $I$ pair and the second $D$ pair are important in the description of the higher-spin states.

Our HF calculation produces a prolate minimum with $\langle \beta \rangle = 0.35$ and $\langle \gamma \rangle = 0^{\circ}$ for $^{112}$Ba and a prolate one with $\langle \beta \rangle = 0.31$ and $\langle \gamma \rangle = 11^{\circ}$ for $^{114}$Ba.
From each of the HF states, we obtain $SDGI$ pairs.
We calculate $^{112}$Ba and $^{114}$Ba using the $SDG_{\rm HF}$- and $SDGI_{\rm HF}$-pair approximations.

The calculated results are presented in Fig. \ref{fig8} and Table \ref{table6}.
The level energies of the low-lying $2^+$, $4^+$, and $6^+$ states obtained by the $SDG_{\rm HF}$ are close to those obtained by the $SDGI_{\rm HF}$, although the latter predicts a slightly larger moment of inertia and slightly larger $B(E2)$ values.

\section{summary}

We have investigated the nucleon-pair approximation of the shell model for medium- and heavy-mass nuclei, using collective nucleon pairs derived by three different approaches:  from the generalized seniority (GS) method, from iterative NPA calculations with the conjugate gradient method (CG), and  from  unconstrained Hartree-Fock calculations (HF).
By dialing between near-spherical and deformed systems in the $pf$, either by changing the relative strength of schematic pairing versus quadrupole-quadrupole interactions or by changing the number of protons,
we found
the NPA with the $SDG_{\rm GS}$ pairs provides good description for nearly-spherical systems, but fails to explain deformed systems, while
the collective feature of deformed systems can be well reproduced by the NPA with the $SDG_{\rm CG}$ pairs.

The conjugate gradient method is computationally very intensive, so we also used Hartree-Fock minima to provide pairs for
transitional nuclei and deformed nuclei. 
We find that the $SDG_{\rm HF}$-pair approximation provides us with good descriptions for low-lying states of the rotational bands.
In particular the $B(E2)$ values obtained by our NPA calculations are very close to the SM results.
The high-spin $I$ pair is responsible for high-spin states of the heavy-mass nuclei, $^{84}$Mo, $^{108-112}$Xe, and $^{112,114}$Ba. In particular we point out our NPA
calculations of low-lying states of $^{112}$Ba and $^{114}$Ba, which are difficult or impossible to be realized in currently large-scale SM calculations due to the huge dimensions of the configuration space.  Thus we demonstrate the utility of
the NPA for nuclear structure physics.

\begin{acknowledgments}
	This material is based upon work supported by
	the National Key R\&D Program of China under Grant No. 2018YFA0404403,
	the U.S. Department of Energy, Office of Science, Office of Nuclear Physics, under Award No. DE-FG02-03ER41272,	
	the National Natural Science Foundation of China under Grant Nos. 12075169 and 11605122,
	the CUSTIPEN (China-U.S. Theory Institute for Physics with Exotic Nuclei) funded by the U.S. Department of Energy, Office of Science grant number DE-SC0009971.
\end{acknowledgments}

\end{document}